# Strained topological insulator spin field effect transistor

**Supriyo Bandyopadhyay**

Department of Electrical and Computer Engineering, Virginia Commonwealth University, Richmond, VA 23284, USA

E-mail: sbandy@vcu.edu

**Abstract**

The notion of a *spin field effect transistor*, where transistor action is realized by manipulating the spin degree of freedom of charge carriers instead of the charge degree of freedom, has captivated researchers for at least three decades. These transistors are typically implemented by modulating the spin orbit interaction in the transistor's channel with a gate voltage, which causes gate-controlled spin precession of the current carriers, and that modulates the channel current flowing between the ferromagnetic source and drain contacts to implement transistor action. Here, we introduce a new concept for a spin field effect transistor which does not exploit spin-orbit interaction. Its channel is made of the conducting surface of a strained three dimensional topological insulator (3D-TI) thin film and the transistor function is elicited by straining the channel region with a gate voltage (using a piezoelectric under-layer) to modify the energy dispersion relation, or the Dirac velocity, of the TI surface states. This rotates the spins of the carriers in the channel and that modulates the current flowing between the ferromagnetic source and drain contacts to realize transistor action. We call it a *strained-topological-insulator-spin-field-effect-transistor*, or STI-SPINFET. Its conductance on/off ratio is too poor to make it useful as a switch, but it may have other uses, such as an extremely energy-efficient stand-alone *single-transistor* frequency multiplier.

## 1. Introduction

The idea of a spin-field-effect-transistor (SPINFET) is at least three decades old and was inspired by the belief that it might take much less gate voltage (and hence much less energy) to manipulate the spins of electrons than to manipulate the number of electrons (charge) in a transistor's channel. Hence, such a transistor could potentially be an extremely low-power device. Early proposals [1] visualized the transistor as having the same structure as a conventional field effect transistor, except its "source" and "drain" contacts are ferromagnetic and act as a *spin polarizer* and a *spin analyzer,* respectively. The source will inject spin polarized electrons into the channel whose spin polarizations would then be controllably rotated while transiting through the channel with a potential applied to the "gate" terminal. That potential varies the spin-orbit interaction in the channel and thus varies the spin rotation. The "drain" contact would filter the arriving spins and preferably transmit those whose polarizations are aligned parallel to its own magnetization, while blocking those that are antiparallel. Therefore, by controlling the spin rotation in the channel with the gate voltage, we can modulate the channel current and realize transistor action. The spin-orbit interaction that is modulated with the gate potential in a conventional SPINFET can be of the Rashba type [1] or the Dresselhaus type [2].

It is now understood that there are three serious challenges with transistors of this variety. First, because spin-orbit interaction is not sufficiently strong in semiconductors, it takes a considerable amount of electric field or potential to induce adequate spin orbit interaction to change the channel current appreciably, meaning that these transistors will have low transconductance and high switching voltages (hence poor energy-efficiency, contrary to expectations) [3]. Second, the ferromagnetic source and drain contacts are typically inefficient spin polarizers and analyzers, so that the modulation of the current (or the conductance on/off ratio) is relatively poor [4], while again contributing to poor transconductance. Third, in *two-dimensional* SPINFETs, ensemble averaging over the transverse wavevector further degrades the conductance on/off ratio. These shortcomings are very difficult to overcome and decades of research have not been able to alleviate them. Nonetheless, the physics of these transistors is intriguing and they may have unusual characteristics, such as an "oscillatory" transfer characteristic, which has other uses.

In this work, we propose and analyze a novel SPINFET. It does not overcome the above shortcomings of SPINFETs, but it employs a very different operating principle, which shows the rich variety of mechanisms that can be called into play to implement SPINFETs. In our device, the gate potential does *not* modulate spin-orbit interaction. In fact, spin-orbit interaction is not needed for transistor action at all. The channel is made of (the surface of) a three dimensional topological insulator (3D-TI) thin film with two (wavevector-dependent) spin eignestates. The ferromagnetic source injects spins with a polarization that is a superposition of the two eigenspin states. The gate voltage mechanically *strains* the TI film, which modulates the Dirac velocity of the surface states, thereby changing the phase relationship in the superposition. That effectively rotates the injected spin, just as the gate voltage rotates the injected spin in the channel of a conventional SPINFET.

The ferromagnetic drain contact acts as a spin analyzer, just as in a conventional SPINFET. When the spin in the channel has been rotated by the gate voltage such that it is parallel to the drain's magnetization when it arrives at the drain contact, it transmits with the highest probability (current is "on"), and if it arrives with spin antiparallel to the drain's magnetization, it transmits with the lowest probability (current is "off"). Thus, transistor action is realized in the same way as the original SPINFET [1, 2], except that here the gate control of the spin rotation is achieved via *strain-induced modulation of the Dirac velocity in the TI surface and not by any modulation of spin-orbit interaction.*

The transistor structure is shown in Fig. 1. The 3D-TI, acting as the transistor's channel, is deposited on a vertically *poled* thin piezoelectric film with two mutually shorted electrodes flanking it, as shown in Fig. 1(b). This shorted pair acts as the *gate*, and this gate configuration is known to generate strain in the intervening region of the piezoelectric, i.e., underneath the 3D-TI film [5]. If we apply a gate voltage whose polarity is such that the resulting (vertical) electric field is directed opposite to the direction of poling, then compressive stress will be generated along the line joining the two electrodes (z-axis in Fig. 1(b)) and tensile stress in the perpendicular direction (x-axis). Reversing the polarity will reverse the signs of the stresses. The use of a piezoelectric thin film deposited on a conducting substrate, as opposed to a piezoelectric substrate, is dictated by the fact that piezoelectrics are insulators and hence a much larger voltage would have been needed to generate a given strain had we substituted the piezoelectric film with a piezoelectric substrate. As long as the piezoelectric film thickness is much larger than the thickness of the TI film, we can assume that 100% of the strain generated in the piezoelectric is transferred to the TI film. The transferred stress/strain changes the energy dispersion relation of the surface states in the 3D-TI, specifically the slope, and hence the Dirac velocity [6]. This results in spin rotation in the TI surface (transistor channel) and that, in turn, modulates the current flowing between the ferromagnetic source and the drain.

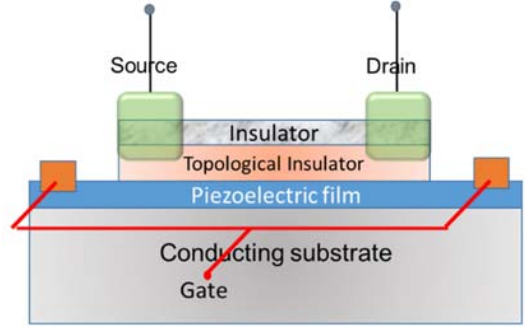

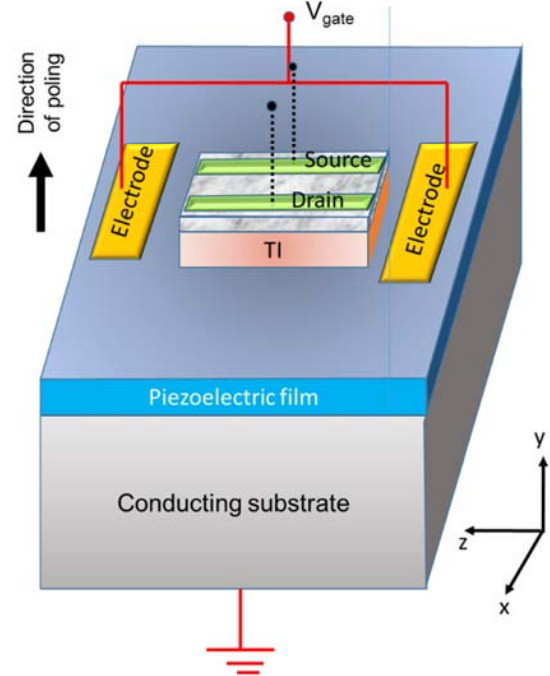

FIG. 1. Structure of a stressed-topological-insulator spin-field-effect-transistor (STI-SPINFET): (a) side view and (b) top view. This diagram is not to scale; the piezoelectric layer is much thicker than the TI layer.

The role of the thin insulating layer between the ferromagnetic source/drain contacts and the TI surface (see Fig. 1) is to act as a tunnel barrier, which is known to improve the spin injection and detection efficiencies of the source/drain contacts [7]. In other words, its presence makes the source and drain contacts better spin polarizer and analyzer. The ferromagnetic contact materials can be those with a high degree of spin polarization, e.g. half metals, to further increase the spin injection/detection efficiency.

Because the device is two-dimensional, ensemble averaging over the transverse wave vector $k_z$ inevitably dilutes the current modulation, very much like the original two-dimensional SPINFET [8], resulting in very poor on/off ratio for the channel conductance. That precludes any use as a switch, but there can be other uses, such as in frequency multiplication, as we discuss later.

We describe the theory of this device in the next section.

## 2. Theory

Fig. 2 shows the conducting surface of the 3D-TI (the channel) pinched between the ferromagnetic source and drain contacts. We can assume that the TI is a common material like $Bi_2Te_3$ or $Bi_2Se_3$. The Hamiltonian describing the surface states near a Dirac point (including higher order terms in the wave vector, up to third order) is [9, 10]

$$H_{TI} = \frac{\hbar^2}{2m^*}\left(k_x^2 + k_z^2\right) + \hbar v_k \left(k_x \sigma_z - k_z \sigma_x\right) + \frac{\lambda}{2}\hbar^3 \left(k_+^3 + k_-^3\right)\sigma_y,$$

where $m^*$ is the effective mass, $\lambda$ is the band warping factor, $v_k = v_0\left(1 + \alpha k^2\right)$, $k^\pm = k_x \pm i k_z$, $v_0$ is the Dirac velocity and the $\sigma$-s are the Pauli spin matrices. To keep the mathematics simple, we will ignore both band warping and the second order correction to the Dirac velocity (i.e., $\alpha = \lambda = 0$), which reduces the Hamiltonian to

$$H_{TI} = \frac{\hbar^2}{2m^*}\left(k_x^2 + k_z^2\right) + \hbar v_0 \left(k_x \sigma_z - k_z \sigma_x\right). \quad (1)$$

We point out that the Hamiltonian in Equation (1) neglects the effect of finite thickness and width, as well as any external magnetic field or spin-orbit interaction.

Diagonalizing the Hamiltonian yields the energy dispersion relation of the spin resolved states as

$$E_\pm = \frac{\hbar^2 k^2}{2m^*} \pm \hbar v_0 k, \quad (2)$$

where $k = \sqrt{k_x^2 + k_z^2}$.

In an "ideal" topological insulator (TI) surface, only the second term in the Hamiltonian in Equation (2) will be present. That will make the energy dispersion relation $E = \pm \hbar v_0 k$, which are the familiar Dirac cones. Real TIs, like $Bi_2Se_3$, however do not fit this bill and the first term in the Hamiltonian will also be present, albeit it will be much smaller than the second term.

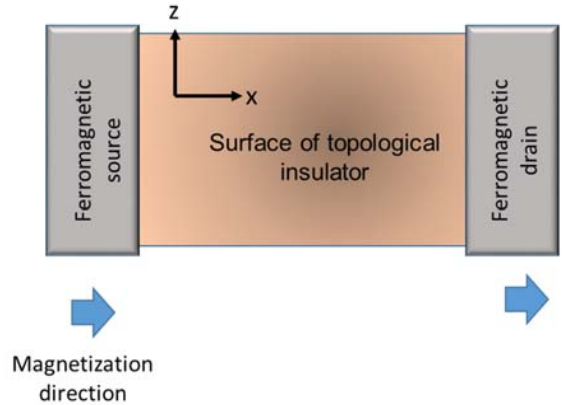

FIG. 2. The surface of the topological insulator film serving as the channel of a transistor with two spin-polarized ferromagnetic contacts (both magnetized in the direction of current flow) acting as the source and the drain

Curiously, with the first term present, the Hamiltonian in Equation (1) looks identical to the Rashba Hamiltonian of a two-dimensional electron gas (2-DEG) with a structural symmetry breaking electric field perpendicular to the 2-DEG (if we replace $\hbar v_0$ with the Rashba constant $\alpha$). The only difference is that in the Rashba system, the first term in the Hamiltonian is dominant over the second term, whereas in the TI, the opposite is true. The Rashba system and the real TI system exhibit similar physics; for example, there is spin-momentum locking in both. The similarities and differences between a real TI and the Rashba system have been discussed in [10].

In Fig. 3, we plot the dispersion relations in Equation (2) assuming $m^* = 0.2 m_0; v_0 = 6.2 \times 10^5$ m/s where $m_0$ is the free electron mass. These values are characteristic of $Bi_2Se_3$.

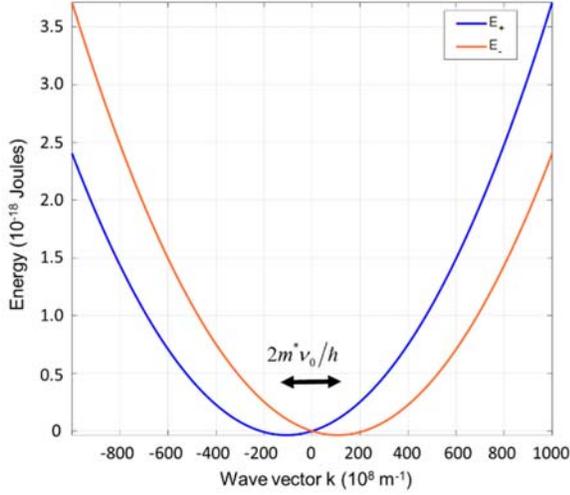

FIG. 3. The energy dispersion relations of the spin split bands on the surface of a TI like $Bi_2Se_3$ or $Bi_2Te_3$

The eigenspinors of the Hamiltonian in Equation (1) are

$$\psi_+ = \begin{bmatrix} \sin\theta \\ \cos\theta \end{bmatrix}; \quad \psi_- = \begin{bmatrix} -\cos\theta \\ \sin\theta \end{bmatrix} \quad (3)$$

where $\theta = (1/2)\arctan(k_z/k_x)$.

We will assume that the TI film is semi-infinite in the z-direction, in which case, the wave vector component $k_z$ is a good quantum number. From Equation (2), it is clear that for any given energy $E$ and magnitude of the wave vector component $k_z$, the magnitudes of the x-components of the wave vectors are *different* in the two spin resolved states. They are related according to

$$E = \frac{\hbar^2 \left[ |k_x^+|^2 + |k_z|^2 \right]}{2m^*} + \hbar v_0 \sqrt{|k_x^+|^2 + |k_z|^2} \quad (4)$$

$$= \frac{\hbar^2 \left[ |k_x^-|^2 + |k_z|^2 \right]}{2m^*} - \hbar v_0 \sqrt{|k_x^-|^2 + |k_z|^2}$$

Hence the angle $\theta$ in Equation (3) is different in the two spin resolved states for any given energy $E$ and $|k_z|$. We will call them $\theta^+$ and $\theta^-$, where $\theta^\pm = (1/2)\arctan(k_z/k_x^\pm)$. Therefore, we should rewrite Equation (3) as $\psi_+ = \begin{bmatrix} \sin\theta^+ \\ \cos\theta^+ \end{bmatrix}$; $\psi_- = \begin{bmatrix} -\cos\theta^- \\ \sin\theta^- \end{bmatrix}$. Note that the inequality between $\theta^+$ and $\theta^-$ is a consequence of the parabolic term in Equation (1) or (2), which is always present (albeit comparatively small) in a real TI. Without that term, $k_x^+ = \pm k_x^-$ and $\theta^+ = \pm\theta^-$ for any given $k_z$.

Let us assume that the ferromagnetic source and the drain contacts are both magnetized in the +x-direction as shown in Fig. 2. We will assume that the source injects only +x-polarized spins into the TI at the complete exclusion of –x polarized spins (perfect spin polarizer). This assumption can be relaxed [8], but it is not necessary at this point for elucidating the principle behind the transistor operation. An injected +x-polarized spin will couple into the two eigenspin states $\psi_+$ and $\psi_-$ in the channel (TI surface) with (wavevector dependent) coupling coefficients $C^+$ and $C^-$. We can view this occurrence as the incident +x-polarized beam splitting into two beams, each corresponding to an eigenspinor in the TI channel. These two beams propagate in different directions since $k_x^+ \neq k_x^-$ for any given energy and $k_z$. Hence the TI channel behaves like a birefringent medium [11]. The beam splitting is expressed by the equation

$$\underbrace{\frac{1}{\sqrt{2}}\begin{bmatrix} 1 \\ 1 \end{bmatrix}}_{+x-polarized} = C^+\psi^+ + C^-\psi^- = C^+\begin{bmatrix} \sin\theta^+ \\ \cos\theta^+ \end{bmatrix} + C^-\begin{bmatrix} -\cos\theta^- \\ \sin\theta^- \end{bmatrix}. \quad (5)$$

The coupling coefficients are found from Equation (5) as

$$C^+ = C^+\left(k_z, k_x^+, k_x^-\right) = \frac{\sin\left(\theta^- + \pi/4\right)}{\cos\left(\theta^+ - \theta^-\right)}$$
$$C^- = C^-\left(k_z, k_x^+, k_x^-\right) = -\frac{\cos\left(\theta^+ + \pi/4\right)}{\cos\left(\theta^+ - \theta^-\right)} \quad (6)$$

In the drain contact, the two beams interfere. The phase difference between them (accrued in traversing the channel) determine the spinor (and hence the spin polarization) of the electron impinging on the drain. This, in turn, determines the transmission probability through the drain contact (spin analyzer) and therefore the source-to-drain current. We will show that the phase difference can be altered with a gate potential which strains the TI and modifies the Dirac velocity, and this elicits the transistor functionality.

The spinor at the drain end is given by

$$[\psi]_{drain} = C^+\begin{bmatrix} \sin\theta^+ \\ \cos\theta^+ \end{bmatrix} e^{i(k_x^+ L + k_z W)} + C^-\begin{bmatrix} -\cos\theta^- \\ \sin\theta^- \end{bmatrix} e^{i(k_x^- L + k_z W)}$$
$$= \frac{\sin(\theta^- + \pi/4)}{\cos(\theta^+ - \theta^-)}\begin{bmatrix} \sin\theta^+ \\ \cos\theta^+ \end{bmatrix} e^{i(k_x^+ L + k_z W)} \quad (7)$$
$$- \frac{\cos(\theta^+ + \pi/4)}{\cos(\theta^+ - \theta^-)}\begin{bmatrix} -\cos\theta^- \\ \sin\theta^- \end{bmatrix} e^{i(k_x^- L + k_z W)}$$

where $L$ is the channel length (distance between source and drain contacts) and *W* is the transverse displacement of the electron as it traverses the channel. Neglecting multiple reflection effects, the transmission amplitude, *t*, is the projection of the arriving spinor on the drain's polarization (which is the +*x*-polarization). Hence

$$t = \frac{1}{\sqrt{2}} e^{ik_z W} \frac{\sin(\theta^- + \pi/4)}{\cos(\theta^+ - \theta^-)} e^{ik_x^+ L} [1 \quad 1] \begin{bmatrix} \sin\theta^+ \\ \cos\theta^+ \end{bmatrix}$$

$$- \frac{1}{\sqrt{2}} e^{ik_z W} \frac{\cos(\theta^+ + \pi/4)}{\cos(\theta^+ - \theta^-)} e^{ik_x^- L} [1 \quad 1] \begin{bmatrix} -\cos\theta^- \\ \sin\theta^- \end{bmatrix}$$

$$= \frac{1}{\sqrt{2}} e^{ik_z W} \frac{\sin(\theta^- + \pi/4)}{\cos(\theta^+ - \theta^-)} e^{ik_x^+ L} (\sin\theta^+ + \cos\theta^+)$$

$$- \frac{1}{\sqrt{2}} e^{ik_z W} \frac{\cos(\theta^+ + \pi/4)}{\cos(\theta^+ - \theta^-)} e^{ik_x^- L} (\sin\theta^- - \cos\theta^-)$$

$$= \frac{e^{ik_z W}}{\cos(\theta^+ - \theta^-)} \sin(\theta^- + \pi/4) \sin(\theta^+ + \pi/4) e^{ik_x^+ L}$$

$$+ \frac{e^{ik_z W}}{\cos(\theta^+ - \theta^-)} \cos(\theta^- + \pi/4) \cos(\theta^+ + \pi/4) e^{ik_x^- L}$$

(8)

The transmission probability *T* is then given by

$$T = |t|^2 = \sin^2(\theta^- + \pi/4)\sin^2(\theta^+ + \pi/4)$$
$$+ \cos^2(\theta^- + \pi/4)\cos^2(\theta^+ + \pi/4) \qquad (9)$$
$$+ (1/2)\cos(2\theta^-)\cos(2\theta^+)\cos\phi$$

where $\phi = (k_x^+ - k_x^-)L$

From Equation (4), we get

$$\sqrt{(k_x^+)^2 + k_z^2} - \sqrt{(k_x^-)^2 + k_z^2} = -2m^* v_0/\hbar, \qquad (10)$$

Defining $k_{av} = \left(\sqrt{(k_x^+)^2 + k_z^2} + \sqrt{(k_x^-)^2 + k_z^2}\right)/2$ and

multiplying both sides of Equation (10) with $2 k_{av}$ yields

$$\phi = (k_x^+ - k_x^-)L = \frac{-2m^* v_0 k_{av} L}{\hbar (k_x^+ + k_x^-)/2} \qquad (11)$$

Note that if $k_z = 0$, which could happen only in a strictly one-dimensional structure, then $\phi = \frac{-2m^* v_0 L}{\hbar}$ which would be independent of the electron wave vector and energy.

Equation (11) shows that we can vary $\phi$ and hence the transmission probability with a gate voltage *if we can change the Dirac velocity $v_0$ with that voltage*. We can do this in the structure shown in Fig. 1. Applying a voltage to the two electrodes in that figure will strain the piezoelectric region pinched between the electrodes. As long as the TI film is much thinner than the piezoelectric film and the insulating film is also thin enough to not clamp the TI, this strain will be transferred almost entirely to the TI film.

In a TI material like Bi$_2$Se$_3$, small stress (or strain) *can change the Dirac velocity $v_0$ along specific crystallographic directions* by $\sim 2 \times 10^4$ m/s per GPa of stress [6]. This provides a handle to vary $\phi$ and hence the transmission probability *T* with an external gate voltage which generates strain in the TI film. That can then modulate the current flowing between the source and the drain contact, thereby realizing transistor action. Since such a transistor consists of three elements – a 3D-TI, strain, and spin interference – we have called it **strained topological insulator spin field effect transistor** (STI-SPINFET).

In the Appendix, we derive an expression for the linear response channel conductance (or source-to-drain conductance) of the STI-SPINFET as a function of the gate voltage to demonstrate the transistor functionality. That expression is

$$G_{SD} = G_0 +$$

$$\frac{q^2 W_z}{2\pi h} \int_0^\infty dE \int dk_z \sqrt{1 - \frac{k_z^2}{k_{av}^2}} \cos\left(\frac{2m^* v_0 L}{\hbar\sqrt{1 - k_z^2/k_{av}^2}}\right) [\delta(E - E_F)]$$

$$= \frac{q^2 W_z}{4\pi h} \int_0^{k_F} dk_z \left\{ \left[1 + \frac{k_z^2}{k_F^2}\right] + 2\sqrt{1 - \frac{k_z^2}{k_F^2}} \cos\left(\frac{2m^* v_0 L}{\hbar\sqrt{1 - k_z^2/k_F^2}}\right) \right\}.$$

(12)

This expression immediately shows that if we can change the Dirac velocity $v_0$ with a gate voltage, then we can change the channel conductance (and hence the source-to-drain current for a fixed drain bias) with the gate voltage, thereby realizing transistor action.

## 2.1. Material considerations

In Bi$_2$Se$_3$, the Dirac velocity in the $\Gamma - \overline{K}$ crystallographic direction is $\sim 6.2 \times 10^5$ m/s under no stress and increases linearly with compressive stress by $\sim 2 \times 10^4$ m/s per GPa [6]. Therefore, this material is a good choice for the TI.

For the piezoelectric layer, one would have preferred to use relaxor materials like PMN-PT because of their high $d_{31}$ (piezoelectric) coefficients, but they are not compatible with TI films. The growth temperature for a TI is typically in the range of $400^0$ – $500^0$ Celsius and the piezoelectric may not survive such high temperatures. There are recent reports of perovskites like (1-*x*)BiScO$_3$-*x*PbTiO$_3$ which can survive temperatures up to $460^0$ C [12] and hence would be compatible with TI growth. It has a $d_{31}$ value of -670 pC/N [12] and therefore is a good choice. With this $d_{31}$ value, one can generate a strain ε of 1000 ppm in the piezoelectric with an electric field $\mathcal{E}$ of 1.5 MV/m which is a very reasonable electric field (ε = $d_{31}\mathcal{E}$).

As mentioned earlier, we will assume that the strain generated in the piezoelectric is completely transferred to the TI. It is a good approximation when the TI film is much thinner than the piezoelectric film. The Young's modulus of $Bi_2Se_3$ nanoribbons was reported as ~40 GPa [13] and we expect it to be about the same in thin films. Hence the stress generated by a strain of 1000 ppm in $Bi_2Se_3$ is 40 MPa. This stress will increase the Dirac velocity $v_0$ in $Bi_2Se_3$ from $6.2\times10^5$ m/s to $6.2\times10^5 +$ 800 m/s, which is enough to modulate the channel conductance of the transistor between the maximum and minimum values. Thus our material choices are $(1-x)BiScO_3-xPbTiO_3$ for the piezoelectric and $Bi_2Se_3$ for the TI.

### 2.2. Transfer characteristics of the transistor

The gate voltage needed to generate a stress $\sigma$ in the TI film is obtained as follows. If we assume that the gate voltage is dropped entirely across the piezoelectric layer (since it is much thicker than the TI layer) and that the conducting substrate has a negligible voltage drop, then we obtain $\sigma = Y\varepsilon = Yd_{31}\mathcal{E} = Yd_{31}V_{gate}/d$, where $d$ is the thickness of the piezoelectric layer and $Y$ is the Young's modulus of the TI. In that case, Equation (12) can be recast as

$$G_{SD} = \frac{q^2 W_z}{4\pi h} \int_0^{k_F} dk_z \left\{ \left[1 + \frac{k_z^2}{k_F^2}\right] + 2\sqrt{1 - \frac{k_z^2}{k_F^2}} \cos\left(\frac{2m^*(\bar{v} + \beta V_{gate})L}{\hbar\sqrt{1 - k_z^2/k_F^2}}\right) \right\},$$

(13)

where $\beta = \zeta d_{31}Y/d$, $\bar{v} = 6.2\times10^5$ m/s (the unstrained Dirac velocity) and $\zeta = 2\times10^{-5}$ m/s per Pa of stress (characteristic of $Bi_2Se_3$ [6]). This expression shows explicit dependence of the transistor's channel conductance on the gate voltage and is hence the transfer characteristic of the transistor.

### 2.3. Numerical results

In Fig. 4, we plot the quantity $G_{SD}/W_z$ as a function of stress from 0 to 40 MPa in steps of 0.5 MPa. The maximum pressure that we consider (40 MPa) is low enough that we can ignore all other pressure-related effects that can show up at extremely high pressures (several GPa). In the upper horizontal axis in Fig. 4, we plot the gate voltage $V_{gate}$ needed to generate the corresponding stress. In this plot, we have assumed that the Fermi wave vector $k_F = 4\times10^{10}$ m$^{-1}$, the effective mass $m^* = 0.2\ m_0$ (where $m_0$ is the free electron mass), $d_{31}$ = -670 pC/N, $d$ = 1 µm, and $L$ = 2000 and 4000 nm. This figure gives the transfer characteristic of the device.

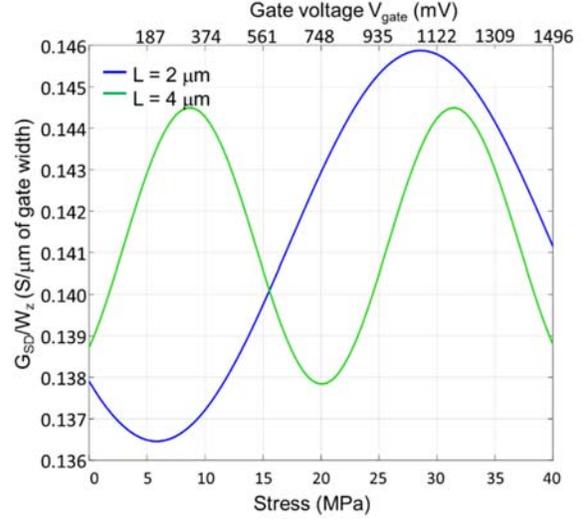

FIG. 4. Transfer characteristic of the strained topological spin field effect transistor shown in green and blue

The disappointing feature in the transfer characteristic is the low conductance on/off ratio which is only about 1.07:1. Imperfect spin injection and filtering at the contacts will reduce this ratio further. Clearly this precludes any application of the STI-SPINFET as a "switch". However, note that the transfer characteristic is *oscillatory*. We can filter out the oscillatory component in the source to drain current with a series capacitor which blocks dc signals. In the gate voltage range 0 to 1.5 V shown in Fig. 4, there are two nearly complete periods of the oscillation in the channel conductance. Hence, if we apply an ac gate voltage with a peak-to-zero amplitude of 1.5 V, the source to drain current (for a fixed drain bias $V_{SD}$) will oscillate with a frequency *four times* that of the gate voltage. This can implement a *frequency multiplier* with a single transistor. In general, the frequency multiplication factor will be $M = 2\left(V_{gate}^{ampl}/V_{period}\right)$, where $V_{gate}^{ampl}$ is the peak-to-zero amplitude of the gate voltage and $V_{period}$ is the period of the oscillation of the source to drain current.

The energy dissipated in the frequency multiplication operation is $C\left(V_{gate}^{ampl}\right)^2 = CM^2 V_{period}^2$ where $C$ is the gate capacitance associated with either of the two electrodes in Fig. 1. The relative dielectric constant $\kappa$ of the perovskite (piezoelectric) is ~2000 [14] and hence the capacitance $C = \kappa\varepsilon_0 A/d =$ 0.18 fF where we have assumed that the electrode area $A$ is 0.1 µm $\times$ 0.1 µm. In the example here, $M$

= 4 and $V_{period}$ = 0.75 V. Therefore, the energy dissipated in the frequency multiplication is ~1.62 fJ, which makes it remarkably energy efficient.

## 3. Conclusions

We have proposed and discussed a new transistor device whose channel is made of a topological insulator (TI) thin film deposited on a piezoelectric film and the source and the drain contacts are ferromagnetic. The piezoelectric is utilized to strain the topological insulator with a gate voltage, which varies the Dirac velocity to rotate spin in the transistor's channel. That allows control of the channel conductance with the gate voltage (because of the spin filtering action of the drain) to implement a transistor. There have been other transistor proposals in the past involving TIs [15-17], but they involve 2D-TIs that are very different from 3D-TIs. A diode based on a 3D-TI has also been proposed [18]. None of them, however, involves "spin" and hence is not a SPINFET.

Unfortunately, the present device cannot be used as a switch because of the extremely poor channel conductance on/off ratio, but the unusual oscillatory transfer characteristic (a hallmark of SPINFETs based on spin interference [1, 2]) can be exploited to implement sub-systems such as a frequency multiplier with a single transistor. Since the sub-system consists of a single device, it has remarkably small footprint and also very low energy dissipation.

## Appendix I: Channel conductance as a function of gate voltage

In calculating the channel conductance as a function of gate voltage, we will ignore self-consistent effects, i.e. we will not invoke the Poisson equation because the surface of a TI is highly conductive. In a highly conductive channel (metallic), any effect of the Poisson equation (such as band bending) will be negligible and hence self-consistency effects can be safely ignored. We will also assume ballistic transport.

The current density in the channel between the source and the drain is given by the Tsu-Esaki formula [19]

$$J_{SD} = \frac{q}{W_y} \int_0^\infty \frac{1}{h} dE \int \frac{dk_z}{\pi} T \left[ f(E) - f(E + qV_{SD}) \right], \quad (A1)$$

where $q$ is the electron's charge, $E$ is the electron (spin carrier) energy, $W_y$ is the thickness of the channel in the y-direction (the vertical extent of the TI surface), $V_{SD}$ is the applied source to drain voltage and $f(\rho)$ is the Fermi-Dirac factor (electron occupation probability) at energy $\rho$ in the source contact. This relation reduces to

$$J_{SD} = \frac{q^2 V_{SD}}{W_y} \int_0^\infty \frac{1}{h} dE \int \frac{dk_z}{\pi} T \left[ -\frac{\partial f(E)}{\partial E} \right] \quad (A2)$$

in the linear response regime when $V_{SD} \to 0$.

The channel conductance is therefore

$$G_{SD} = \frac{I_{SD}}{V_{SD}} = \frac{J_{SD} W_y W_z}{V_{SD}} = \frac{q^2 W_z}{\pi h} \int_0^\infty dE \int dk_Z T \left[ -\frac{\partial f(E)}{\partial E} \right]. \quad (A3)$$

Substituting the expression for $T$ in Equation (9) into Equation (A3), we obtain

$$G_{SD} = G_0 + \frac{q^2 W_z}{2\pi h} \int_0^\infty dE \int dk_z \cos(2\theta^-) \cos(2\theta^+) \cos\phi \left[ -\frac{\partial f}{\partial E} \right] \quad (A4)$$

where

$$G_0 = \frac{q^2 W_z}{4\pi h} \int_0^\infty dE \int dk_z \left[ 1 + \sin(2\theta^+) \sin(2\theta^-) \right] \left[ -\frac{\partial f}{\partial E} \right]$$

$$= \frac{q^2 W_z}{4\pi h} \int_0^\infty dE \int dk_z \left[ 1 + \frac{k_z}{\sqrt{(k_x^+)^2 + k_z^2}} \frac{k_z}{\sqrt{(k_x^-)^2 + k_z^2}} \right] \left[ -\frac{\partial f}{\partial E} \right] \quad (A5)$$

$$\approx \frac{q^2 W_z}{4\pi h} \int_0^\infty dE \int dk_z \left[ 1 + \frac{k_z^2}{k_{av}^2(E)} \right] \left[ -\frac{\partial f}{\partial E} \right] \text{ if } |k_x^+ - k_x^-| \ll k_{av}.$$

At low temperatures, the quantity $G_o$ becomes

$$G_0 \approx \frac{q^2 W_z}{4\pi h} \int_0^\infty dE \int dk_z \left[ 1 + \frac{k_z^2}{k_{av}^2(E)} \right] \left[ \delta(E - E_F) \right],$$

$$= \frac{q^2 W_z}{4\pi h} \int_0^{k_F} dk_z \left[ 1 + \frac{k_z^2}{k_F^2} \right] \quad (A6)$$

where $k_F$ is the Fermi wave vector (the wave vector at the Fermi energy $E_F$).

Equation (A4) can be re-written as

$$G_{SD} = G_0 + \frac{q^2 W_z}{2\pi h} \int_0^\infty dE \int dk_z \frac{k_x^-}{\sqrt{(k_x^-)^2 + k_z^2}} \frac{k_x^+}{\sqrt{(k_x^+)^2 + k_z^2}} \cos\phi \left[ -\frac{\partial f}{\partial E} \right]$$

(A7)

Again, if $|k_x^+ - k_x^-| \ll k_{av}$, then Equation (11) can be recast as $\phi \approx \frac{-2m^* v_0 k_{av} L}{\hbar \sqrt{k_{av}^2 - k_z^2}}$ and thereafter Equation (A7) can be re-written as

$$G_{SD} = G_0 +$$

$$\frac{q^2 W_z}{2\pi h} \int_0^\infty dE \int dk_z \sqrt{1 - \frac{k_z^2}{k_{av}^2}} \cos\left( \frac{2m^* v_0 L}{\hbar \sqrt{1 - k_z^2/k_{av}^2}} \right) \left[ -\frac{\partial f(E)}{\partial E} \right],$$

which, at low temperatures, reduces to

$$G_{SD} = G_0 + \frac{q^2 W_z}{2\pi h} \int_0^\infty dE \int dk_z \sqrt{1 - \frac{k_z^2}{k_{av}^2}} \cos\left(\frac{2m^* v_0 L}{\hbar \sqrt{1 - k_z^2/k_{av}^2}}\right) \left[\delta(E - E_F)\right]$$

$$= \frac{q^2 W_z}{4\pi h} \int_0^{k_F} dk_z \left\{ \left[1 + \frac{k_z^2}{k_F^2}\right] + 2\sqrt{1 - \frac{k_z^2}{k_F^2}} \cos\left(\frac{2m^* v_0 L}{\hbar \sqrt{1 - k_z^2/k_F^2}}\right) \right\}.$$

(A9)

which is Equation (12).

*Appendix II: The STI-SPINFET with an ideal 3D-TI surface as the channel*

Let us consider a hypothetical "ideal" 3D-TI surface where the parabolic term in the Hamiltonian [see Equation (2)] is absent. This will change Equation (4) to

$$E = \hbar v_0 \sqrt{|k_x^+|^2 + |k_z|^2} = -\hbar v_0 \sqrt{|k_x^-|^2 + |k_z|^2}.$$ (A10)

Squaring the above equation immediately yields

$$k_x^+ = \pm k_x^-; \quad \theta^+ = \pm \theta^-.$$ (A11)

The first case $(k_x^+ = k_x^-; \; \theta^+ = \theta^-)$ is trivially uninteresting since then $\phi = 0$ always and there is no conductance modulation and hence no transistor action. The second case $(k_x^+ = -k_x^-; \; \theta^+ = -\theta^-)$ is interesting and allows for transistor action. *However, this case corresponds to interference between a forward propagating state and a reverse propagating state.*

Considering the second case, the transmission amplitude [Equation (8)] will change to

$$t = \frac{e^{ik_z W}}{\cos(2\theta^+)} \sin(-\theta^+ + \pi/4) \sin(\theta^+ + \pi/4) e^{ik_x^+ L}$$
$$+ \frac{e^{ik_z W}}{\cos(2\theta^+)} \cos(-\theta^+ + \pi/4) \cos(\theta^+ + \pi/4) e^{-ik_x^+ L} \quad \text{(A12)}$$
$$= \frac{1}{2}\left[1 + e^{-i\varphi}\right].$$

where $\varphi = 2k_x^+ L$. This yields the transmission probability as

$$T = |t|^2 = \cos^2(\varphi/2).$$ (A13)

In this case, the transmission probability can vary from 0 (complete reflection) to 1 (complete transmission) at any given energy and transverse wavevector.

From Equation (A10), we will get that

$$k_x^+ = \sqrt{\left(\frac{E}{\hbar v_0}\right)^2 - k_z^2}; \quad \varphi = 2L\sqrt{\left(\frac{E}{\hbar v_0}\right)^2 - k_z^2}$$

(A14)

$$T = \cos^2\left(L\sqrt{\left(\frac{E}{\hbar v_0}\right)^2 - k_z^2}\right)$$

Finally, Equation (A3) will change to

$$G_{SD} = \frac{I_{SD}}{V_{SD}} = \frac{J_{SD} W_y W_z}{V_{SD}}$$

$$= \frac{q^2 W_z}{\pi h} \int_0^\infty dE \int dk_Z \cos^2\left(L\sqrt{\left(\frac{E}{\hbar v_0}\right)^2 - k_z^2}\right) \left[-\frac{\partial f(E)}{\partial E}\right].$$

(A15)

Once again, by changing $v_0$ with a gate voltage, we can change the source-to-drain channel conductance and realize transistor action. At low temperatures, Equation (A15) will simplify to

$$G_{SD} = \frac{q^2 W_z}{\pi h} \int_0^{k_F} dk_z \cos^2\left(L\sqrt{\left(\frac{E_F}{\hbar v_0}\right)^2 - k_z^2}\right).$$ (A16)

**Acknowledgement**
The author is grateful to Ms. Raisa Fabiha and Ms. Rahnuma Rahman for the plots in Figs. 3 and 4 and to Prof. Avik W. Ghosh for discussions on the effect of stress on the Dirac velocity in topological insulators.

**Conflict of Interest:** The author declares no conflict of interest.

**Data Availability Statement**: No new data were generated in this work.